\documentclass[fleqn,12pt]{wlscirep}
\usepackage[utf8]{inputenc}
\usepackage[T1]{fontenc}
\usepackage{xcolor,color,soul}
\usepackage{subcaption}
\usepackage{lineno}
\usepackage{xr}
\makeatletter
\newcommand*{\addFileDependency}[1]{
  \typeout{(#1)}
  \@addtofilelist{#1}
  \IfFileExists{#1}{}{\typeout{No file #1.}}
}
\makeatother

\newcommand*{\myexternaldocument}[1]{%
    \externaldocument{#1}%
    \addFileDependency{#1.tex}%
    \addFileDependency{#1.aux}%
}
\myexternaldocument{supporting-info}

\usepackage{hyperref}
\hypersetup{
    colorlinks,
    linkcolor={blue!90!black},
    citecolor={magenta!80!black},
    urlcolor={blue!90!black}
}
\title{Edge-Induced Excitations in Bi$_2$Te$_3$ from Spatially-Resolved Electron Energy-Gain Spectroscopy}
\author[1,$\dag$]{Helena La}
\author[1,$\dag$]{Abel Brokkelkamp}
\author[1,$\dag$]{Stijn van der Lippe}
\author[2,3]{Jaco ter Hoeve}
\author[2,3]{Juan Rojo}
\author[1,*]{Sonia Conesa-Boj}
\affil[1]{Kavli Institute of Nanoscience, Delft University of Technology, Delft, 2628 CJ, The Netherlands}
\affil[2]{Nikhef Theory Group, Science Park 105, 1098 XG Amsterdam, The Netherlands}
\affil[3]{Department of Physics and Astronomy, VU, 1081 HV Amsterdam, The Netherlands}

\affil[$\dag$]{Equal contribution}
\affil[*]{Corresponding author: s.conesaboj@tudelft.nl}


\begin{abstract}
Among the many potential applications
of topological insulator materials, their broad potential for the development of novel tunable plasmonics at THz and mid-infrared frequencies for quantum computing, terahertz detectors, and spintronic devices is particularly attractive. 
The required understanding of the intricate relationship between nanoscale crystal structure and the properties of 
the resulting plasmonic resonances remains, however,
elusive for these materials.
Specifically, edge- and surface-induced
plasmonic resonances, and other collective
excitations, are often buried beneath the continuum
of electronic transitions, making it difficult to isolate and interpret these
signals using  techniques such as electron energy-loss spectroscopy (EELS). 
Here we focus on the experimentally clean
energy-gain EELS region to characterise 
collective excitations in 
the topologically insulating material
Bi$_2$Te$_3$ 
and correlate them with the underlying
crystalline structure with nanoscale
resolution.
We identify with high significance
the presence of a distinct energy-gain peak around $-0.8$ eV, with spatially-resolved maps
revealing that its  intensity is markedly enhanced at the edge regions of the specimen.
Our findings illustrate the reach 
of  energy-gain EELS analyses
to accurately map collective excitations in quantum materials, a key asset
in the quest towards new tunable plasmonic devices.

\end{abstract}
\begin{document}

\flushbottom
\maketitle
\newcommand{\bite}{\text{Bi}_2 \text{Te}_3}
\newcommand{\todo}{\hl{\textbf{TODO}}}

\thispagestyle{empty}

\section*{Introduction}
Topological insulator (TI) materials,
such as $\bite$~\cite{Zhang2009, Xia2009, Hsieh2009} and Bi$_2$Se$_3$,
possess unique properties that make them well suited for the design of nanoplasmonic devices operating in the THz and mid-infrared frequency ranges.~\cite{Grigorenko2012,DiPietro2013,Marta2015,Yin2017,giorgianni2016strong}  
Topological insulators can also support  plasmonic excitations,
collective oscillations of electrons that interact strongly with light or other electrons and lead
to enhanced light-matter interactions such as strong scattering, absorption, and emission.
In particular, low-energy plasmons~\cite{Zhao2015}
have been reported in $\bite$ below $3$ eV while correlated plasmons 
at energies $\sim 1$ eV have been identified for Bi$_2$Se$_3$~\cite{Whitcher2020}. 
In this context, advancing our understanding of how to optimally deploy TIs for the development of tunable plasmonic devices that operate efficiently in optical frequencies has the potential to benefit a wide range of applications, including quantum computing~\cite{Kitaev2006, Fu2008}, terahertz detectors~\cite{Zhang2010}, and spintronic devices~\cite{Chen2009}.

In recent years, significant progress has been achieved in resolving plasmon resonances at the nanoscale, providing valuable information about the spatial and spectral distribution of plasmonic modes.
To this end, electron-based
spectroscopic techniques such as electron energy-loss spectroscopy (EELS) have 
demonstrated their suitability to investigate the electronic and optical properties of a wide
range of materials, including the study of their plasmonic resonances.~\cite{Chu2009,Garcia2010,Myroshnychenko2012,Coenen2015,Nerl2017,Lagos2017,Yankovich2019}
In parallel, advances in transmission electron microscopy (TEM) have resulted in
novel opportunities for scrutinizing the functionalities of nanostructured materials. 
For instance, the incorporation of monochromators and aberration correctors makes it possible
to resolve collective lattice oscillations (phonons)
and study them with nanometer spatial resolution.~\cite{Krivanek2014,Egoavil2014,Govyadinov2017,Venkatraman2019,Hage2020,Qi2021} 
%
Furthermore, the incorporation of machine learning (ML) algorithms for EELS data analysis and interpretation has further enhanced
the reach of spectroscopic techniques
to pin down the properties of nanomaterials.
As recently demonstrated~\cite{Brokkelkamp2022,Roest2021}, ML methods enable
the spatially-resolved determination of local electronic properties such as the band gap and the dielectric function with nanometer resolution from EELS spectral images

Here we investigate low-energy collective excitations in 
the TI material Bi$_2$Te$_3$ 
by means of EELS spectral images focusing on the energy-gain ($\Delta E< 0$) region.~\cite{Boersch1966,Asenjo2013,Lagos2017,Idrobo2018}
As compared to traditional EELS, this strategy offers the key advantage that gain peaks are not
obscured by the multiple scatterings continuum
and other electronic transitions
taking place in the energy-loss region
($\Delta E> 0$), 
enabling the clean identification of narrow collective
excitations with enhanced  spectral resolution.
The resulting characterisation of Bi$_2$Te$_3$ specimens
makes it possible to search for collective resonances in the low-gain region,
and correlate their spatial distribution with
distinct structural features such as surfaces, edges, and regions with sharp
thickness variations.

Our analysis reveals the presence of a narrow energy-gain peak around $-0.8$ eV 
whose intensity is the largest in regions of the specimen
associated with exposed edges and surfaces.
We demonstrate the robustness of our energy-gain peak identification algorithm
with respect to the strategy adopted for the modelling and subtraction of the
dominant zero-loss peak (ZLP) background,
quantify the statistical significance of this
signal,
and estimate procedural uncertainties by means
of the Monte Carlo method widely used in high-energy
physics.
Our work represents a significant 
step forward in exploiting the information
contained in the energy-gain region
of EELS spectral images
to achieve an improved understanding
of localised collective resonances in  TI materials.

\section*{Results and Discussion}
\begin{figure}[t!]
    \centering
    \includegraphics[width=\textwidth]{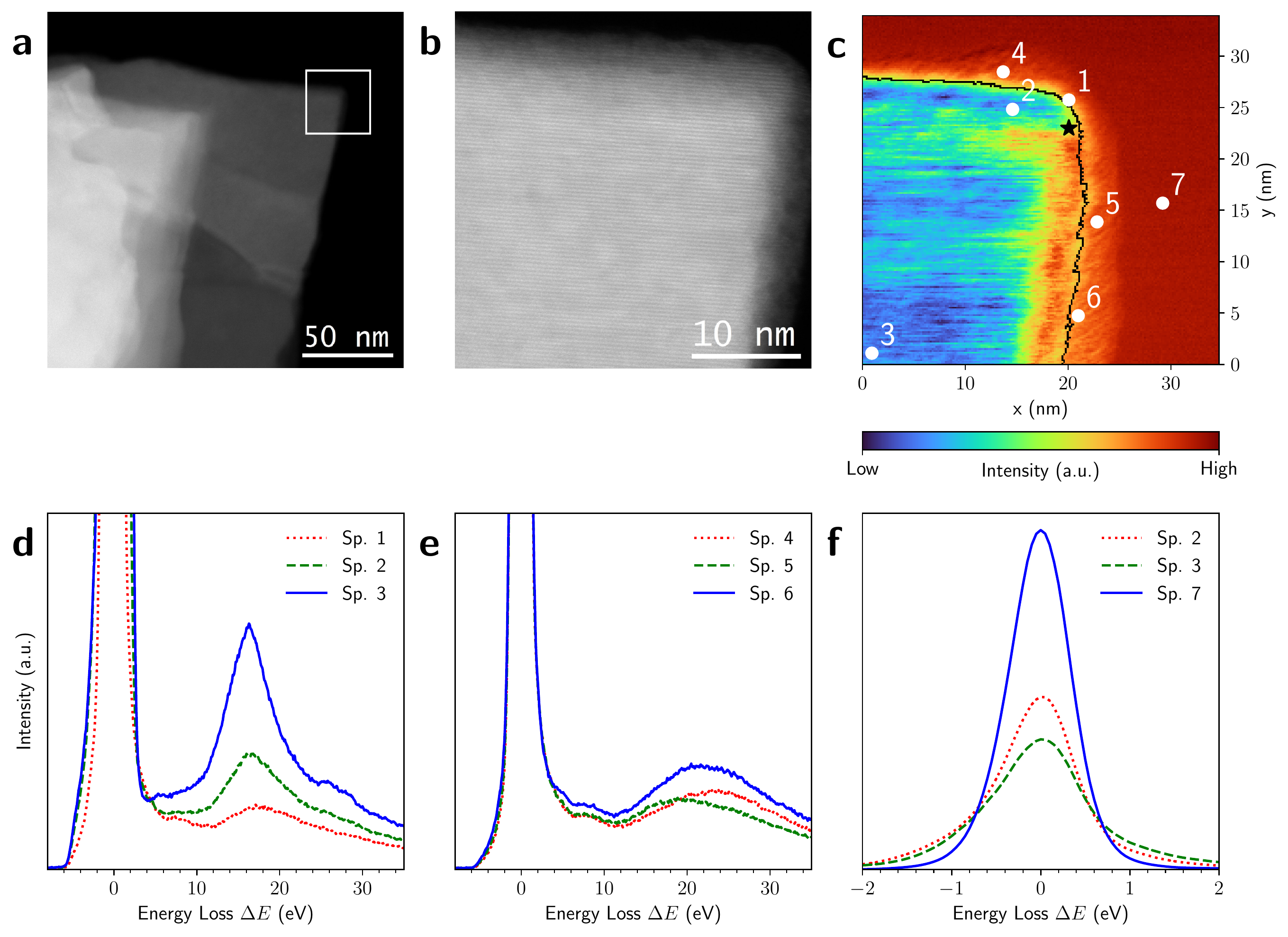}
    \caption{\textbf{Spatially-resolved EELS analysis of $\bite$.}
    \textbf{(a)} HAADF-STEM image of a representative $\bite$ flake. 
    \textbf{(b)} Magnified region of the top right corner of the specimen around 
    the white square in \textbf{(a)}.
    \textbf{(c)} EELS spectral image corresponding
    to the white square region in \textbf{(a)}.
    The color map corresponds to the total
    integrated intensity in each pixel.
    The black line indicates the edge of the $\bite$ specimen, determined from the thickness map as described in the text.
    \textbf{(d)} EELS spectra 
    corresponding to the different regions
    of the specimen indicated in \textbf{(c)}: between the vacuum and the edge (sp. $1$); the vicinity of the edge towards the inner region (sp. $2$); and  the innermost, thicker part (sp. $3$). 
    Spectra sp. $1$, sp. $2$ and sp. $3$ 
    display the  bulk plasmon peak of $\bite$ at around $16$~eV.
    Additionally, sp. $3$ also shows the Bi $\text{O}_{4,5}$ edges excited from Bi 5$d$ electrons at $25.6$ eV and $27.9$ eV.
    \textbf{(e)} Same as \textbf{(d)} for 
    EELS spectra in the immediate vicinity of the $\bite$ edge,  displaying characteristic features of $\text{Bi}_2 \text{O}_3$. 
    \textbf{(f)} A comparison of sp. 2, sp. 3, and sp. 7 in the low loss and gain regions ($|\Delta E| \le 2$ eV).}
    \label{fig:HAADF_spectra}
\end{figure}

Fig.~\ref{fig:HAADF_spectra}\textbf{a} displays a high-angle annular dark-field (HAADF) Scanning Transmission Electron Microscopy (STEM) image of a representative $\bite$ specimen. 
For closer examination, Fig.~\ref{fig:HAADF_spectra}\textbf{b} shows the magnified top right corner of the same specimen.
Further characterisation of the atomic structure 
of this specimen is provided in SI-\ref{sec:atomic_structure_characterisation} of the Supplementary Material.
By means of electron energy-loss spectroscopy (EELS), we acquire a spectral image (Fig.~\ref{fig:HAADF_spectra}\textbf{c})
of the specimen
in the same region, indicated 
with a white square  in Fig.~\ref{fig:HAADF_spectra}\textbf{a}.
The color map corresponds to
the total integrated intensity in each pixel. 
The black line indicates the edge of the $\bite$ specimen, which is automatically determined
from the spatially-resolved thickness map associated to the spectral image~\cite{Brokkelkamp2022},
specifically from its local rate of change.

Fig.~\ref{fig:HAADF_spectra}\textbf{d} displays EELS spectra taken at three different locations within the spectral image, labelled as spectra sp$1$, sp$2$, and sp$3$ in the following.
Spectra sp$1$, sp$2$, and sp$3$ are acquired in the region between the vacuum and the edge of the specimen, in the vicinity of the specimen edge towards the inner region, and in the innermost part of the specimen, respectively. 
The three spectra reveal the presence of distinct spectral features located at approximately energy losses of $8.6$ eV and $16.6$ eV, where the latter corresponds to the bulk plasmon peak in accordance with previous studies.~\cite{Peranio2011}
Furthermore, the peaks at $25.6$ eV and $27.9$ eV 
observed in sp$3$ can be identified with the Bi O$_{4,5}$ edges excited from Bi 5$d$ electrons, also reported in the literature.~\cite{nascimento1999}
Fig.~\ref{fig:HAADF_spectra}\textbf{e} compares
three other EEL spectra (labelled as sp$4$, sp$5$, and sp$6$) acquired in the immediate vicinity of the specimen edge.
The three spectra exhibit a broad peak located around $21$ eV, which can be identified with the bulk plasmon of $\text{Bi}_2\text{O}_3$.~\cite{Pau2017,Li2021} 
It is worth nothing that the presence of $\text{Bi}_2\text{O}_3$ in the surfaces
of the specimen is is not visible from the HAADF images.
The reason is that HAADF
intensity scales with $Z^n$, with $Z$ being the atomic number, which is much smaller in O as compared to Te.

Fig.~\ref{fig:HAADF_spectra}\textbf{f}
compares the EELS intensities in the region
of energy losses $\Delta E$ restricted
to the window $\left[ -2~{\rm eV},2~{\rm eV}\right]$
for spectra sp$2$, sp$3$, and sp$7$.
This comparison illustrates the dependence of the 
dominant Zero-Loss Peak (ZLP) background with respect to the location in the specimen:
bulk (sp$3$), close to edge (sp$2$), and vacuum (sp$7$).
On the one hand, as one moves from the vacuum towards the bulk region, the ZLP intensity gradually decreases.
This effect can be ascribed to the greater number of inelastic scattering events that occur in the bulk (thicker) regions, compared to the vacuum where the beam electrons do not experience inelastic scatterings.
On the other hand, we also observe an enhanced intensity in the specimen regions as compared
to the vacuum for $|\Delta E| \ge 0.6$ eV,
highlighting material-sensitive contributions
to the spectra
which contain direct information
on its local electronic properties.

Removing this ZLP background is instrumental in
order to identify the presence of localised collective
excitations such as phonons~\cite{Govyadinov2017} and plasmon peaks~\cite{Yankovich2019} 
in the low energy-loss region.
The same considerations apply to the cleaner energy-gain region~\cite{Idrobo2018}, where the continuum of inelastic scattering contributions is absent.
Here we model the ZLP in terms of a Gaussian
distribution following the procedure
described in SI-\ref{sec:subtraction-procedure},
with the fitting region
restricted to $[-0.4, 0.4]$ eV to remove
the overlap with $\Delta E$ values 
at which plasmonic modes of $\bite{}$
have been reported.~\cite{Zhao2015}
Subsequently, the ZLP is removed pixel by pixel
in the EELS spectral image
and the resulting spectra are 
inspected to identify peaks and other well-defined features
in an automated manner.
We note that the small band gap~\cite{Chen2009, Hsieh2009} of $\bite{}$, $E_{\rm bg}\sim 0.15$ eV, prevents reliably training deep learning models
for the ZLP parametrisation and subtraction
as done in previous studies from our group.~\cite{Brokkelkamp2022,Roest2021,vanheijst2021}
Furthermore, although here we focus on a $\bite{}$ 
specimen, the procedure is fully general and 
applicable to other materials which can be inspected
with EELS.

\begin{figure}[htbp]
    \centering
    \includegraphics[width=.95\textwidth]{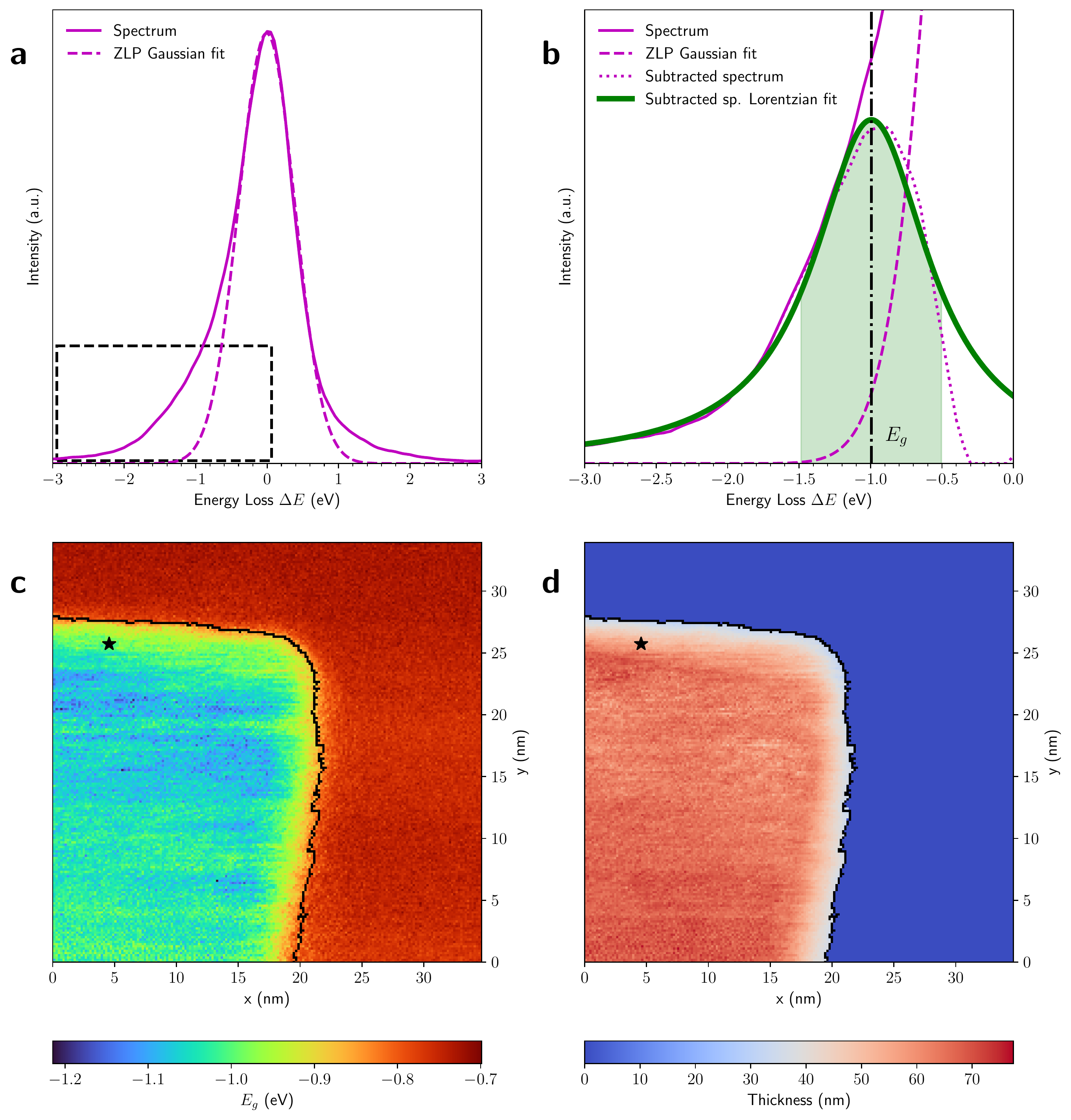}
    \caption{\textbf{Energy-gain peak identification in $\bite$.} 
    \textbf{(a)} The EEL spectrum (solid) in the pixel indicated 
    with a star in
Fig.~\ref{fig:HAADF_spectra}\textbf{c}
    together with the corresponding fit
    to the ZLP  (dashed curve).
    \textbf{(b)} Close-up of the dashed rectangle in \textbf{(a)}, now adding  the Lorentzian fit (thick solid line) to the subtracted spectrum (dotted line).
    The vertical line indicates the
    mean of the 
    Lorentzian energy-gain peak $E_g$,
    while the filled region indicates 
    the corresponding FWHM.
    \textbf{(c)} Spatially-resolved map
    displaying the location
    of the energy-gain peak $E_g$,
    determined following
    the procedure of \textbf{(b)} across
    the whole spectral image
    of Fig.~\ref{fig:HAADF_spectra}\textbf{c}.
    \textbf{(d)} Same as \textbf{(c)}
    now for the local specimen thickness.
    The dark blue region beyond the specimen edge corresponds to the vacuum region of the spectral image.
    }
    \label{fig:subtraction_procedure}
\end{figure}

Fig.~\ref{fig:subtraction_procedure} summarises the adopted strategy for the spatially-resolved identification of energy-gain peaks. 
First, Fig.~\ref{fig:subtraction_procedure}\textbf{a} 
shows the EEL spectrum  for the pixel indicated 
with a star in
Fig.~\ref{fig:HAADF_spectra}\textbf{c} together with the corresponding ZLP fit.
Closing up on the energy-gain region, 
Fig.~\ref{fig:subtraction_procedure}\textbf{b}
displays the resulting subtracted spectrum,
to which a Lorentzian function is fitted
 (see SI-\ref{sec:subtraction-procedure}
for details) to  extract the position $E_g$
and intensity  of the dominant energy-gain peak. 
The procedure is repeated for the 
complete EELS spectral image, making it possible
to construct the spatially-resolved map
of $E_g$ shown
in Fig.~\ref{fig:subtraction_procedure}\textbf{c} across the inspected
region of the $\bite{}$ specimen.
As in Fig.~\ref{fig:HAADF_spectra}\textbf{c}, the
black line indicates the boundary of the $\bite{}$ sample.
Fig.~\ref{fig:subtraction_procedure}\textbf{c} 
reveals the presence of an energy-gain
peak in the specimen with $E_g$ values
between $-1.1$ eV and $-0.85$ eV.
We demonstrate in SI-\ref{sec:subtraction-procedure} that results
for the ZLP removal and energy-gain peak
identification are robust with respect
to the choice of ZLP model function.
Then Fig.~\ref{fig:subtraction_procedure}\textbf{d}
displays the $\bite{}$ thickness
map as obtained from the deconvolution of the single-scattering EELS distribution~\cite{Brokkelkamp2022}.
The dark blue region beyond the
specimen corresponds to either the vacuum
or the $\text{Bi}_2\text{O}_3$ regions.
In the edge region there is a sharp
increase in thickness, while in the bulk
region the thickness exhibits an approximately
constant value of 70 nm. 

\begin{figure}[htbp]
    \centering
    \includegraphics[width=.95\textwidth]{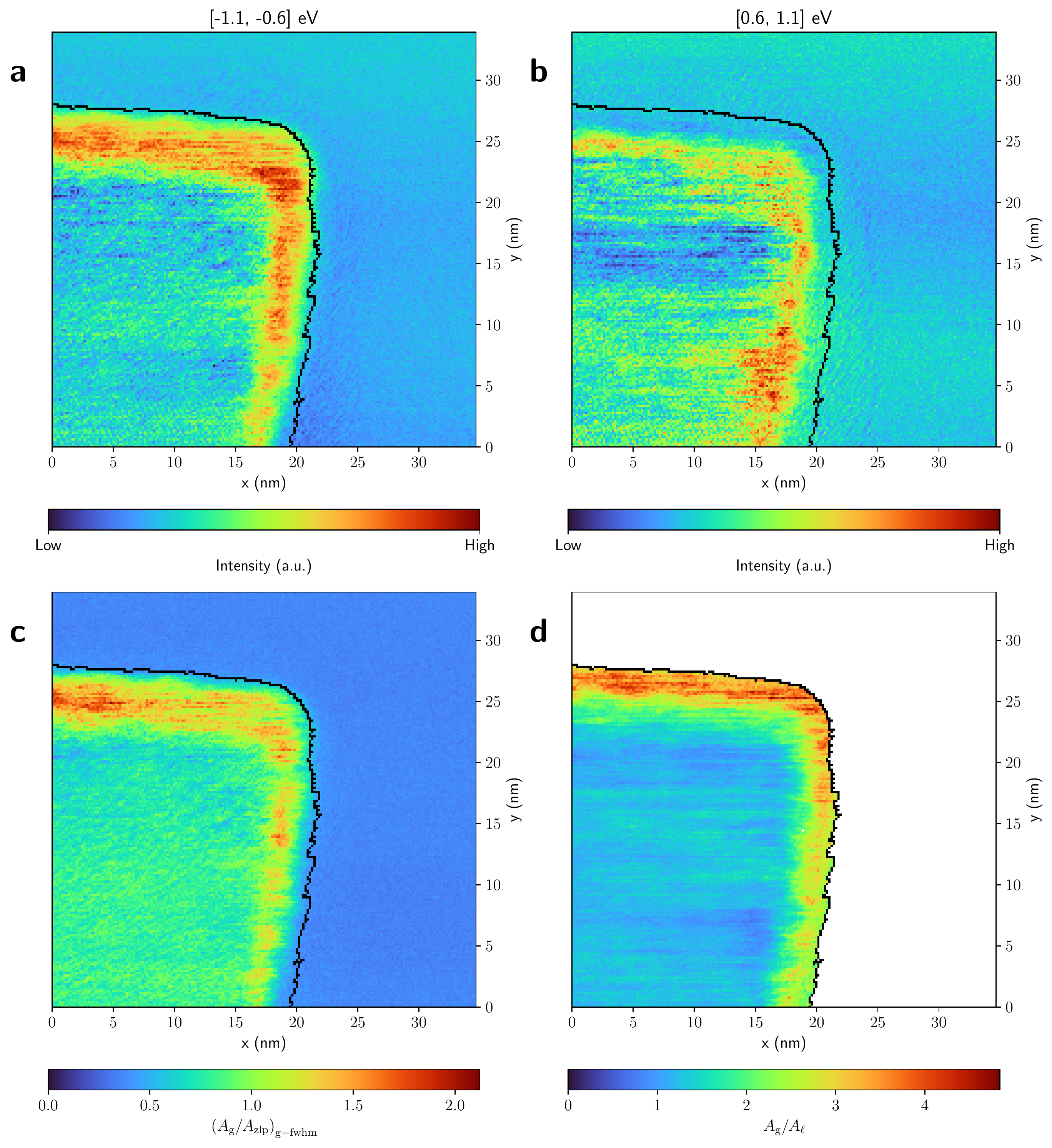}
    \caption{\textbf{Spatially-resolved characterisation of energy-gain peaks.} 
    \textbf{(a)} 
    Same as Fig.~\ref{fig:HAADF_spectra}\textbf{c}, now with the intensity of the 
    EEL spectra (after ZLP subtraction) 
    integrated 
    in the window $[-1.1, -0.6]$ eV
    where the gain peak
    identified in Fig.~\ref{fig:subtraction_procedure}\textbf{c} is located.
    \textbf{(b)} Same as \textbf{(a)}
    for the mirrored energy-loss window,
    $[0.6, 1.1]$ eV.
    \textbf{(c)} Spatially-resolved
    map of the ratio $s_g$, Eq.~(\ref{eq:sg_ratio}), defined as
    the area under the FWHM of the
    Lorentzian fit to the energy-gain peak,
    filled region in Fig.~\ref{fig:subtraction_procedure}\textbf{b},
    to the area under the ZLP in the same
    $\Delta E$ window.
    \textbf{(d)} Ratio of the area $A_g$ under the FWHM of the Lorentzian fit to the energy-gain peak to 
    its counterpart $A_\ell$ in the loss region,
    where the  vacuum region is masked out for clarity.
    }
    \label{fig:Bi2Te3-gain_maps_area_ratio}
\end{figure}

In order to further characterise the energy-gain
peak identified in
Fig.~\ref{fig:subtraction_procedure}\textbf{c} 
and to correlate its properties with
local structural features of the specimen,
Figs.~\ref{fig:Bi2Te3-gain_maps_area_ratio}\textbf{a} and \textbf{b} 
display the intensity of the
ZLP-subtracted EEL spectra integrated
in the energy windows
$\left[ -1.1,-0.6\right]$ eV
and $\left[ 0.6,1.1\right]$ eV 
for the gain and loss regions respectively. 
These $\Delta E$ intervals are chosen 
to contain the range of $E_g$ values
displayed in Fig.~\ref{fig:subtraction_procedure}\textbf{c} 
and then mirrored to the energy-loss region.
In the latter case,
the  EEL spectra receive additional contributions
to the inelastic scattering distribution beyond those considered here.
The most notable feature of Fig.~\ref{fig:Bi2Te3-gain_maps_area_ratio}\textbf{a}
is an enhancement of the integrated intensity
in the edge region of the specimen characterised
by a sharp variation of the local thickness
(Fig.~\ref{fig:subtraction_procedure}\textbf{d}).
 
To quantify the statistical significance of
the identified energy-gain peak, it
is convenient to evaluate the ratio 
\begin{equation}
s_{g} \equiv \left( A_{g} / A_{\rm zlp}\right)_{\rm g-fwhm} \, ,
\label{eq:sg_ratio}
\end{equation}
where $A_{g}$ and $A_{\rm zlp}$ are defined as the areas
under the full width at half-maximum (FWMH) of the Lorentzian fit signal, filled region in Fig.~\ref{fig:subtraction_procedure}\textbf{b},
and under the ZLP in the same $\Delta E$ region,
respectively.
In other words,  $s_g$ measures the significance
of the energy-gain peak in units of the 
ZLP background.
Fig.~\ref{fig:Bi2Te3-gain_maps_area_ratio}\textbf{c} displays a spatially-resolved map of $s_g$ across the specimen.
The region of enhanced intensity reported
in Fig.~\ref{fig:Bi2Te3-gain_maps_area_ratio}\textbf{a} and associated
to the specimen edge 
corresponds to the highest values of $s_g$
in Fig.~\ref{fig:Bi2Te3-gain_maps_area_ratio}\textbf{c},
reaching up to a factor two.
This high significance
confirms that the observe intensity
enhancement in the gain region 
is a genuine feature of the data
rather than an artifact of the ZLP removal procedure.

It is also interesting to compare the features
of the approximately symmetric
 peaks appearing in the energy-gain and energy-loss regions, whose values $E_g$ and $E_\ell$ 
 respectively are mapped across the specimen
 in Fig.~\ref{fig:peak_location} of the Supplementary Material. 
One observes in general
a stronger intensity of the energy-gain
  peak as compared to its energy-loss counterpart.
 To quantify this observation and to compare
 their relative intensities, we display
 in Fig.~\ref{fig:Bi2Te3-gain_maps_area_ratio}\textbf{d} the ratio $A_g / A_\ell$ of the areas under the FWHM of the energy-gain Lorentzian fit to that of the energy-loss peak. 
 The vacuum region is masked out to facilitate
 readability.
 As can be seen, in the bulk of the sample the ratio 
 $A_g / A_\ell$ is of the order unity, whereas in the edge region
 of the specimen the ratio   reaches a factor of around 4.
 The latter result indicates that
 surface and edge effects
 enhance the  relative intensity
 of the energy-gain peak.
 The combination pf Figs.~\ref{fig:subtraction_procedure} and~\ref{fig:Bi2Te3-gain_maps_area_ratio} demonstrates
 the presence of a well-defined, significant energy-gain peak in $\bite{}$ located 
 around $E_g\simeq -0.9$ eV whose intensity
 is enhanced in the edge regions of the specimen
 close to the boundary.

\begin{figure}[htbp]
    \centering
    \includegraphics[width=.95\textwidth]{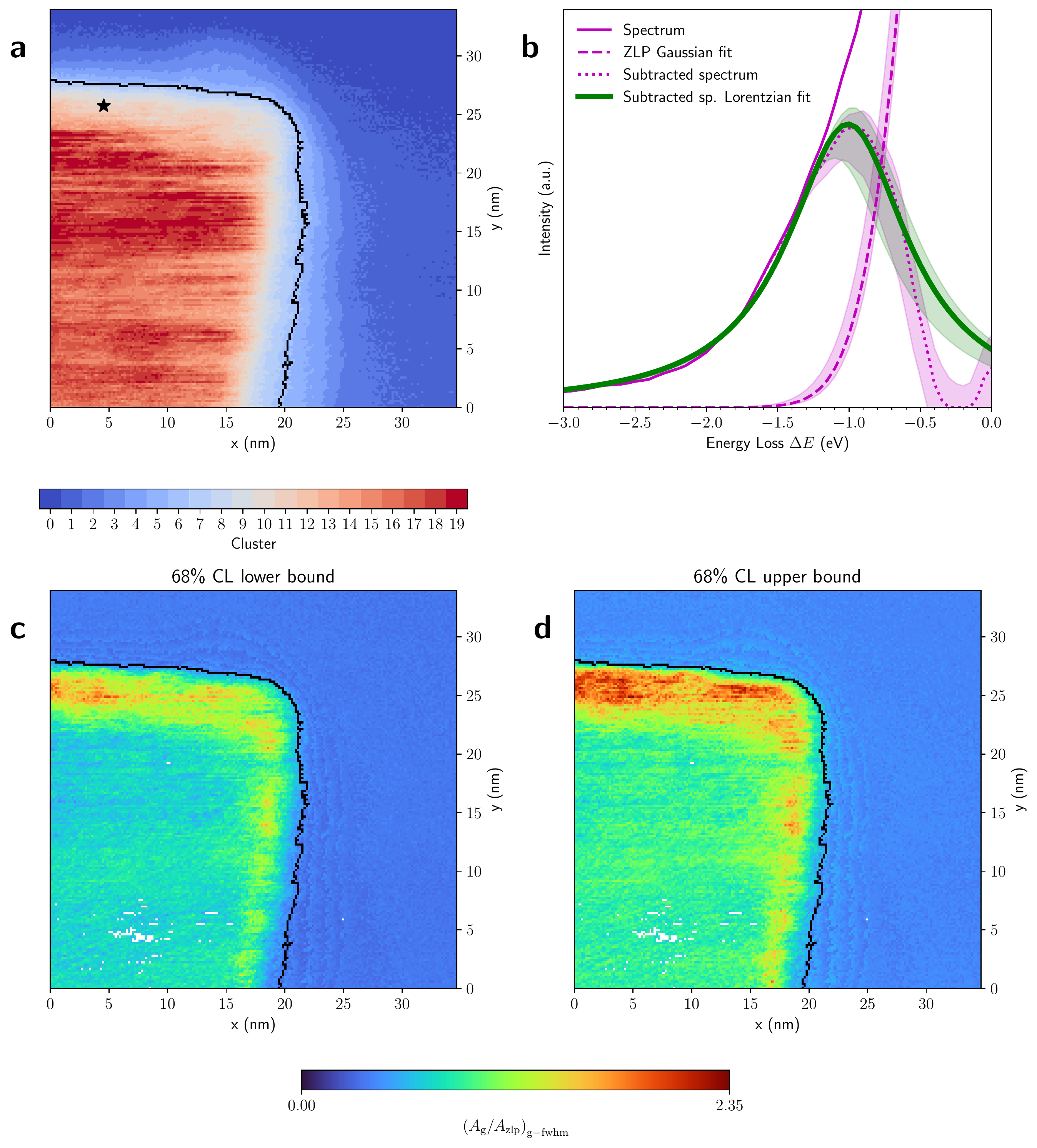}
    \caption{\textbf{Energy-gain peak
    characterisation with the Monte Carlo replica method.} \textbf{(a)} 
    The EELS spectral image of Fig.~\ref{fig:HAADF_spectra}\textbf{c}
    classified into 20 clusters, each
    of them composed by pixels with
    similar thickness.
    \textbf{(b)} 
    Same as Fig.~\ref{fig:subtraction_procedure}\textbf{b}
    now using the Monte Carlo replica method
    to estimate and propagate the ZLP
    fitting model uncertainties. 
    For the ZLP Gaussian fit, the subtracted spectrum, and the Lorentzian fit to the latter
    we display both the median over replicas
    and the 68\% CL intervals.
    \textbf{(c,d)} Same as Fig.~\ref{fig:Bi2Te3-gain_maps_area_ratio}\textbf{c}
    now the lower and upper ranges, 
    respectively, of the
    68\% CL interval for the area ratio
    evaluated over the Monte Carlo replicas.
    See Fig.~\ref{fig:median_and_bounds} in the Supplementary Material for the corresponding median map.
    }
    \label{fig:area_ratio_uncertainty}
\end{figure}

 A potential limitation of this
 analysis concerns the lack of a systematic
 estimate of the functional uncertainties
 associated to the ZLP modelling and its
 subsequent subtraction from the EELS spectral image.
 To this purpose, we deploy the Monte Carlo replica method for error propagation, originally
 developed for proton structure studies
 in high-energy physics~\cite{Watt2012, TheNNPDFCollaboration2007,NNPDF:2014otw,Hartland:2019bjb,Candido:2023utz} and then extended
 to deep learning models of the ZLP
 within the {\sc\small EELSfitter} framework.~\cite{Brokkelkamp2022,Roest2021} 
 First, one applies $K$-means clustering to the EELS spectral image  with the similarity
 measure being the area under the three
  bins of the EELS intensity around $\Delta E=0$,
  which operates as
 a proxy for the local thickness map of Fig.~\ref{fig:subtraction_procedure}\textbf{d}.
 This procedure 
 results in the 20 clusters 
 shown in Fig.~\ref{fig:area_ratio_uncertainty}\textbf{a},
 each of them composed by pixels with similar
thickness.
Within each cluster, the EELS intensities are assumed to be sampled from the same underlying distribution,
and $N_{\mathrm{rep}}$ spectra (``replicas'')
are randomly selected from each cluster.
By fitting a separate ZLP model to each
replica, one ends up with a sampling of $N_{\mathrm{rep}}$ models of the ZLP which
can be used to estimate 
uncertainties and propagate them to
the subtracted spectra and the subsequent
Lorentzian fits. 

Fig.~\ref{fig:area_ratio_uncertainty}\textbf{b}
displays the same ZLP-substracted spectrum 
as in Fig.~\ref{fig:subtraction_procedure}\textbf{b}  now with the Monte Carlo replica method used 
to estimate  ZLP model uncertainties.
For the ZLP  fit, the subtracted spectrum, and the Lorentzian fit to the latter 
the bands indicate the 68\% confidence level
(CL) intervals evaluated over the $N_{\rm rep}$
replicas. 
By repeating this approach in all clusters, we calculate the area ratio $s_g$ defined
in Eq.~(\ref{eq:sg_ratio}) for all pixels in the spectral image using the replicas to propagate uncertainties. 
This results in lower and upper bounds of the 68\% confidence interval of the area ratio shown in Fig.~\ref{fig:area_ratio_uncertainty}\textbf{c} and \textbf{d} respectively.
The corresponding map of the  median of $s_g$
is consistent with that reported in  Fig.~\ref{fig:Bi2Te3-gain_maps_area_ratio}\textbf{c} and shown in Fig.~\ref{fig:median_and_bounds} in the Supplementary Material. 
Given that a good significance (above unity) of the energy-gain peak is still observed in the map of 
the lower limit of the 68\% CL interval for the 
relevant edge region, one can 
conclude that the results of this work
are not distorted by unaccounted-for
methodological or procedural
uncertainties.

To confirm the reproducibility of our findings, we have performed additional measurements on a different  $\bite$ specimen characterised by the same crystal structure and with comparable features as the one discussed here.
The resulting analysis is summarised in SI-\ref{sec:other_specimen} of the Supplementary Material and reveals the same qualitative features
in the energy-gain region, namely a well-defined,
narrow peak at energy gains around $-0.7$ eV
whose intensity is enhanced in edge and surface regions and whose significance reaches values of $s_g\sim 4$. 
This independent analysis further confirms the robustness of our results, in particular the strong
correlation between the enhanced intensity
of the energy-gain peak located around
$[-0.9, -0.7]$ eV and specimen regions
displayed sharp thickness variations including
edges and surfaces.

It is beyond our scope to identify the
underlying physical phenomena leading to the observed
edge- and surface-induced energy-gain peaks
in $\bite{}$.
Several mechanisms have been explored
leading to resonance signatures
in the $\Delta E$ region relevant for
our results, such as
wedge Dyakonov waves~\cite{Talebi2016} and
edge- and surface-located Dirac-plasmons
in the closely related TI material Bi$_2$Se$_3$.
One can in any case exclude thermal effects associated
to a Bose-Einstein distribution, given
that states with $\sim 1 $ eV have a very low 
occupation probability at room temperatures.
Disentangling the specific mechanisms explaining 
our observations requires dedicated
theoretical simulations mapping the EELS response
of $\bite{}$ with different structural and geometric
configurations and is left for future work.

\section*{Summary and Outlook}

In this work we have presented a systematic,
spatially-resolved
investigation of the energy-gain region
of EELS spectral images acquired on Bi$_2$Te$_3$ 
specimens.
The main motivation was to avoid the inelastic
continuum that pollutes the energy-loss region,
which may prevent identifying exotic
phenomena appearing at $\Delta E$ values below
a few eV. 
An automated peak-identification procedure identifies a narrow feature located around $\Delta E\sim -0.8$ eV whose intensity and significance
are strongly enhanced in regions
characterised by sharp
thickness variations, such as surfaces and edges. 
We assess the role of methodological uncertainties
associated to e.g. the ZLP subtraction procedure
and find that our results are robust against
them. 
The observed resonance could be the signature of edge- and surface-plasmons such as those reported in Bi$_2$Se$_3$, thought dedicated simulations would be required to unambiguously ascertain its origin.

While here we focus in $\bite{}$ as a proof-of-concept,
our approach for ZLP substraction and energy-gain peak tracking is fully general and can be deployed to any specimen for which EELS-SI measurements are acquired, and
in particular it is amenable to atomically thin materials of the van der Waals family.
Our approach is made available
in the new release of the {\sc\small EELSfitter}
framework and hence can be straightforwardly
used by other researchers aiming to
explore the information
contained in the energy-gain region of EELS-SI
to identify, model, and correlate localised collective excitations in nanostructured materials.
Possible future improvements include the extension
to multiple gain-peaks deconvolution and the
improved modelling of the loss region describing the
inelastic continuum background.
All in all, our findings illustrate the powerful reach 
of  energy-gain EELS 
to accurately map and characterise the signatures of collective excitations and other exotic resonances arising in quantum materials.

\section*{Methods}

\textbf{Specimen preparation.}
The specimen used in this study were $\bite$ flakes that were mechanically exfoliated from bulk crystals through sonication in isopropanol (IPA) at a ratio of $2$ mg of $\bite$ per $1$ ml of IPA. 
The exfoliated flakes were then transferred onto holey carbon grids for EELS investigations. \\[-0.3cm]

\noindent
\textbf{STEM-EELS settings.}
The scanning transmission electron microscopy (STEM) images and electron energy-loss/gain spectra were obtained using a JEOL200F monochromated equipped with aberration corrector and a Gatan Imaging Filter (GIF) continuum spectrometer. The instrument was operated at $200$ kV and the convergence semi-angle was $14$ mrad. The collection semi-angle for EELS acquisition was $18.3$ mrad obtained by inserting a $5$ mm EELS entrance aperture. The EELS dispersion was $50$ meV per channel. \\[-0.3cm]

\noindent
\textbf{Data processing and interpretation.}
The spatially-resolved maps of the energy-gain peaks and the associated peak identification, fitting, and data analysis techniques were performed using the open-source Python package {\sc\small EELSfitter}.
All features presented in this work
are available in its latest public release
together with the accompanying input EELS spectral images via its {\sc\small GitHub} repository.

\section*{Declaration of Competing Interest}

The authors declare that they have no known competing financial interests or personal relationships that could have appeared to influence the work reported in this paper.

\section*{Funding}

H.~L., A.~B., and S.C.-B. acknowledge financial support from ERC through the Starting Grant “TESLA” grant agreement no. 805021. 
The work of J.~R. is partially supported by NWO (Dutch Research Council)
and by an ASDI (Accelerating Scientific Discoveries) grant from the Netherlands eScience Center.
The work of J.~t.~H is 
supported by NWO (Dutch Research Council).

\section*{Supplementary Material}

The supplementary material of this manuscript provides technical details on the atomic structure characterisation of $\bite{}$, the energy-gain peak identification procedure, the uncertainty estimate using the Monte Carlo replica method,
and the analysis of the energy-gain peaks in different $\bite{}$ specimens.

\clearpage


\begin{center}
  {\LARGE \textbf{\Huge{Supporting Information:}}}\\[0.9cm]

      {\Large \textbf{Edge-Induced Excitations in Bi$_2$Te$_3$ from Spatially-Resolved\\[0.3cm] Electron Energy-Gain Spectroscopy}}\\[0.9cm]
  
  {Helena La,$^{1}$ Abel Brokkelkamp,$^{1}$ Stijn van der Lippe,$^{1}$ Jaco ter Hoeve,$^{2,3}$ \\Juan Rojo,$^{2,3}$ Sonia Conesa-Boj$^{1,*}$}\\[0.5cm]

    \small{
    $^1$Kavli Institute of Nanoscience, Delft University of Technology, 2628 CJ, Delft, The Netherlands\\
    $^2$Nikhef Theory Group, Science Park 105, 1098 XG Amsterdam, The Netherlands\\
    $^3$Department of Physics and Astronomy, VU, 1081 HV Amsterdam, The Netherlands\\
    $^*$Corresponding author. Email: s.c.conesaboj@tudelft.nl 
  }
\end{center}

\tableofcontents
\clearpage

\section{Atomic structure characterisation of $\bite{}$}\label{sec:atomic_structure_characterisation}

\begin{figure}[htbp]
\centering
    \includegraphics[width=.8\textwidth]{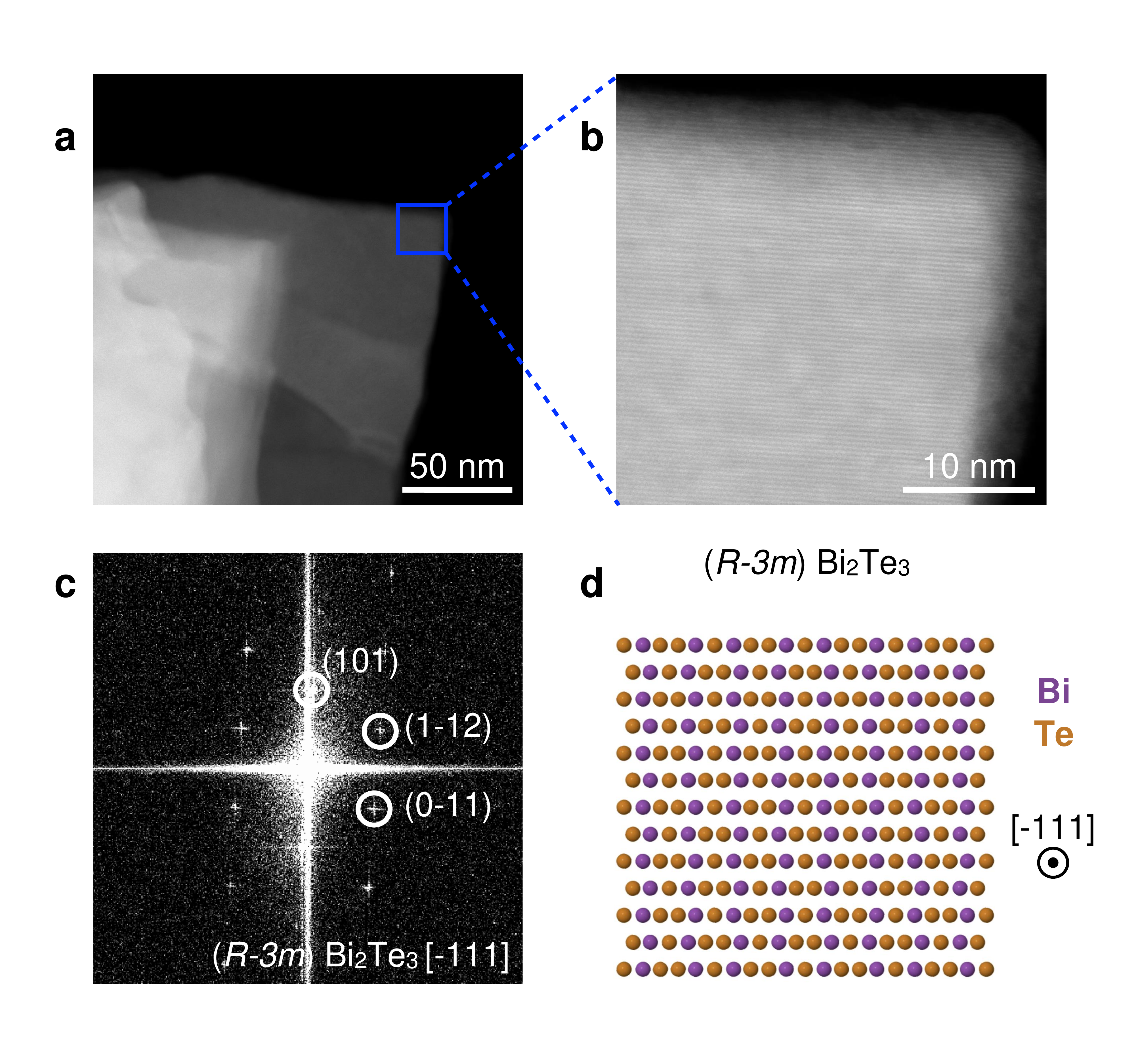}
    \caption{\textbf{STEM analysis of the $\bite{}$ crystal analysed.} \textbf{(a)} Low-magnification
      HAADF-STEM image 
    of the $\bite{}$ specimen
    considered in this work.
    \textbf{(b)} Atomic-resolution HAADF-STEM image of the region indicated with a blue squared in \textbf{(a)}. The $\bite{}$ crystal is oriented along the $[-1~1~1]$ direction, as shown in the corresponding Fast Fourier transform (FFT) in \textbf{(c)}. \textbf{(d)} Atomic model of the $\bite{}$ crystal viewed along the 
     $[-1~1~1]$ direction.}
    \label{fig:Bi2Te3-haadf-stem}
\end{figure}

Fig.~\ref{fig:Bi2Te3-haadf-stem}
presents the STEM analysis of the $\bite{}$ specimen
considered in the main manuscript.
Fig.~\ref{fig:Bi2Te3-haadf-stem}\textbf{a} shows a low-magnification high-angle annular dark-field (HAADF) STEM image taken on the same $\bite{}$ crystal
inspected by EELS. 
The corresponding atomic-resolution HAADF-STEM image of the region indicated with a blue square in Fig.~\ref{fig:Bi2Te3-haadf-stem}\textbf{a} is 
displayed in Fig.~\ref{fig:Bi2Te3-haadf-stem}\textbf{b}.
As well known, $\bite{}$ is characterised by 
a hexagonal primitive cell.
From the corresponding fast Fourier transform (FFT), Fig.~\ref{fig:Bi2Te3-haadf-stem}\textbf{c}, 
it is found that this $\bite{}$ crystal is oriented along the $[-1~1~1]$ direction.
Fig.~\ref{fig:Bi2Te3-haadf-stem}\textbf{d} shows an atomic model of the $\bite{}$ crystal viewed along the same $[-1~1~1]$ direction, displaying
the characteristic hexagonal
crystal structure of $\bite{}$.

\clearpage

\section{Peak identification procedure}
\label{sec:subtraction-procedure}

Here we describe the procedure used to 
identify and characterise energy-gain peaks in $\bite{}$. 
We demonstrate how our approach is
robust with respect to the model adopted
for the ZLP.
As mentioned in the main manuscript, while
here we focus on a $\bite{}$ specimen, 
our energy-gain peak identification algorithm is 
general and could be applied to 
other materials.

The results presented in the main
manuscript are based on a Gaussian
model for the ZLP,
    \begin{equation}
    \label{eq:gaussian_ZLP}
        I_{\rm ZLP}(\Delta E) 
            = 
            I_{\rm ZLP}^{(\rm max)} \exp{\left( -\frac{(\Delta E - \Delta E_0)^2}{2\sigma^2} \right)} \, ,
    \end{equation}
 with $\Delta E$ being the energy
loss (and hence $\Delta E<0$ implies
energy gains), $\Delta E_0$ and $\sigma^2$
being the mean and variance of
    the distribution,
    and $I^{(\rm max)}_{\rm ZLP}$ is the maximal
    intensity of the ZLP.
This Gaussian model for the ZLP,
Eq.~(\ref{eq:gaussian_ZLP}), is
fitted independently to each individual
spectra composing the spectral image.
The fit region is restricted
to $[-0.4, 0.4]$ eV in order to remove
the possible overlap with those
values of $\Delta E$
at which plasmonic modes of $\bite{}$
have been previously reported.
Subsequently, the ZLP Gaussian model Eq.~(\ref{eq:gaussian_ZLP}) is removed
from the measured spectral image and the resulting subtracted spectra are
inspected to
identify peaks or other relevant features
in a fully automated manner.

In this work we identify energy-gain peaks
by fitting the subtracted spectra in the energy region $[-3.5, -0.4]$ eV with a Lorentzian function given by
    \begin{equation}
    \label{eq:gaussian_Lorentzian}
        I_\text{peak} (\Delta E)
            = 
            I_{\max} \,\frac{\Gamma^2}{\Gamma^2 + (\Delta E - E_g)^2}
    \end{equation}
where $I_{\max}$ is the height of the peak,
$\Gamma$ its width,
and $E_g$ its median,
the latter corresponding to the most likely location
of the peak being identified.
Once one has determined the
parameters of the Lorentzian
model Eq.~(\ref{eq:gaussian_Lorentzian})
for each of the pixels composing
the spectral image,
we can determine relevant estimators
such as the FWHM and the area underneath it.
This way, it is possible to determine
the statistical significance of
the observed features.
Although in this work
we consider only single-peak
identification, 
in the presence of multiple
features in the energy-gain region
Eq.~(\ref{eq:gaussian_Lorentzian}) could
be extended to a sum of a series of independent Lorentzian 
resonances.

The robustness of the results
presented in this work upon variations
of the model assumed for the ZLP is demonstrated
by repeating the analysis using
other functional forms for the ZLP model.
Specifically, in addition
to the Gaussian model Eq.~(\ref{eq:gaussian_ZLP})
we consider 
a split Gaussian function, 
a Pearson VII function, and 
a Pseudo-Voigt function to
parameterise the ZLP, all 
other aspects of the fitting procedure
unchanged.
These three models are 
described
below and have been considered
in the EELS literature in the context
of the ZLP subtraction.

\paragraph{Split Gaussian function.}
This model is the same as 
the Gaussian function of  Eq.~(\ref{eq:gaussian_ZLP}) with the difference that the variance is different at the left side
and at the right side of the peak.
An asymmetric model such a split Gaussian
may be more adequate to describe
the ZLP in the presence of an intrinsic
asymmetry in the energy loss $\Delta E$.

\paragraph{Pearson VII function.} This function is often used to describe peak shapes from $X$-ray powder diffraction patterns~\cite{PearsonVII} and is defined as
    \begin{equation}
        I(\Delta E)=I_{\max } \frac{w^{2 m}}{\left[w^2+\left(2^{1 / m}-1\right)\left(\Delta E-\Delta E_0\right)^2\right]^m} \, ,
    \end{equation}
    with $w$ defining the peak width.
    The model parameter $m$
    can be adjusted and here we find
    that $m=4$ provides the best 
    description of the ZLP.
    
\paragraph{The pseudo-Voigt function.}
The Voigt function is defined
as the convolution of two broadening functions, one a Gaussian and the
other a Lorentzian function.
The pseudo-Voigt function is also a popular choice in describing peak shapes in diffraction patterns~\cite{Pseudo-Voigt} and approximates the exact convolution by a linear combination of a Gaussian and a Lorentzian.
Their mixing can be adjusted by a
model parameter $\eta$ as follows:
    \begin{equation}
        I(\Delta E)=I_{\max } \left[ \eta G \left(\Delta E, \sigma^2 \right) + \left(1 - \eta \right) L\left(\Delta E, \sigma^2 \right) \right],
    \end{equation}
    where $G \left(\Delta E, \sigma^2 \right)$
    and $L\left(\Delta E, \sigma^2 \right) $ are normalised Gaussian and Lorentzian functions
    with a common variance $\sigma^2$ respectively. 
    
\begin{figure}[htbp]
    \centering
    \includegraphics[width=0.99\textwidth]{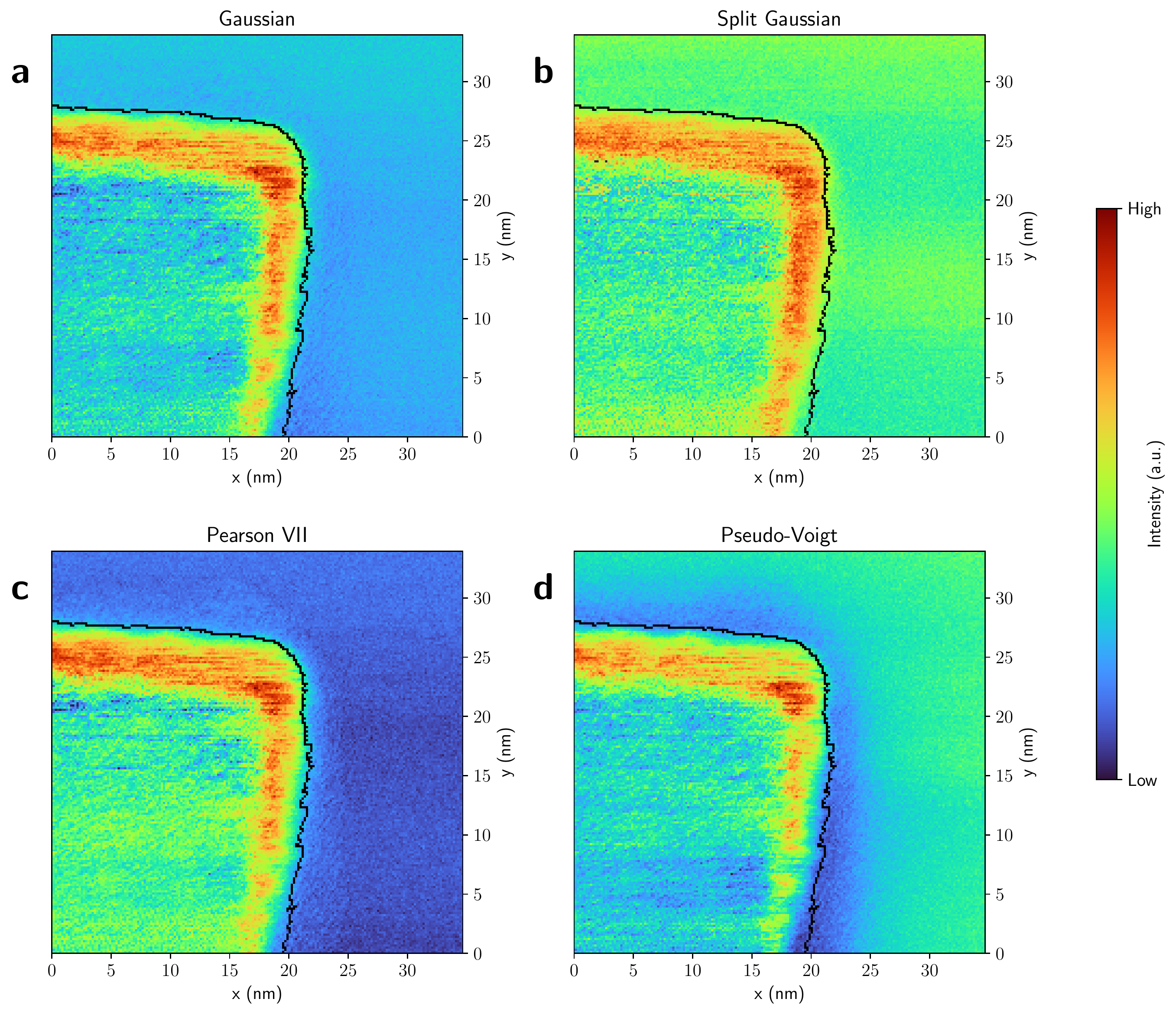}
    \caption{Same as Fig.~\ref{fig:Bi2Te3-gain_maps_area_ratio}\textbf{a} in the main manuscript, 
    displaying the ZLP-subtracted EELS intensity
    integrated in the region $[-1.1,-0.6]$ eV
    around the identified energy-gain peak.
    We compare the outcome of our
    baseline choice,
    a Gaussian model for the ZLP \textbf{(a)},
    with the corresponding results
    based on \textbf{(b)} split Gaussian, \textbf{(c)} Pearson VII, and \textbf{(d)} pseudo-Voigt functions for the ZLP
    parametrisation. 
    Each panel adopts its own intensity range,
    and similar colors in the different panels in general do not correspond to similar intensity values.
    Irrespective of the choice of ZLP model function, the highest integrated intensities
    are found near the  edge region indicated by a black curve.
    }
    \label{fig:ZLP_model_comparison}
\end{figure}

\paragraph{Stability upon choice of ZLP model.}
Fig. \ref{fig:ZLP_model_comparison}
displays, as done in Fig.~\ref{fig:Bi2Te3-gain_maps_area_ratio}\textbf{a} of the main manuscript, the ZLP-subtracted EELS intensity integrated in the region $[-1.1,-0.6]$ eV
around the identified energy-gain peak.
        We compare the outcome of our
    baseline choice,
    a Gaussian model for the ZLP \textbf{(a)},
    with the corresponding results
    based on \textbf{(b)} split Gaussian, \textbf{(c)} Pearson VII, and \textbf{(d)} pseudo-Voigt functions for the ZLP
    parametrisation. 
    Each panel adopts its own intensity range,
    and similar colors in the different panels in general do not correspond to similar intensity values.
    Irrespective of the choice
    of ZLP model function, the highest
    integrated intensity is found
    in the edge region of the specimen.
    We have verified that
    other relevant properties
    of the energy-gain peak at
    $\Delta E\simeq -1$ eV, such
    as its mean and width,
    are also stable with respect to this choice.
    We conclude that our energy-gain peak identification algorithm is robust with respect
    to the modelling of the ZLP.

\clearpage

\section{Peak location in the energy-loss region}

\begin{figure}[htbp]
    \centering
    \includegraphics[width=.99\textwidth]{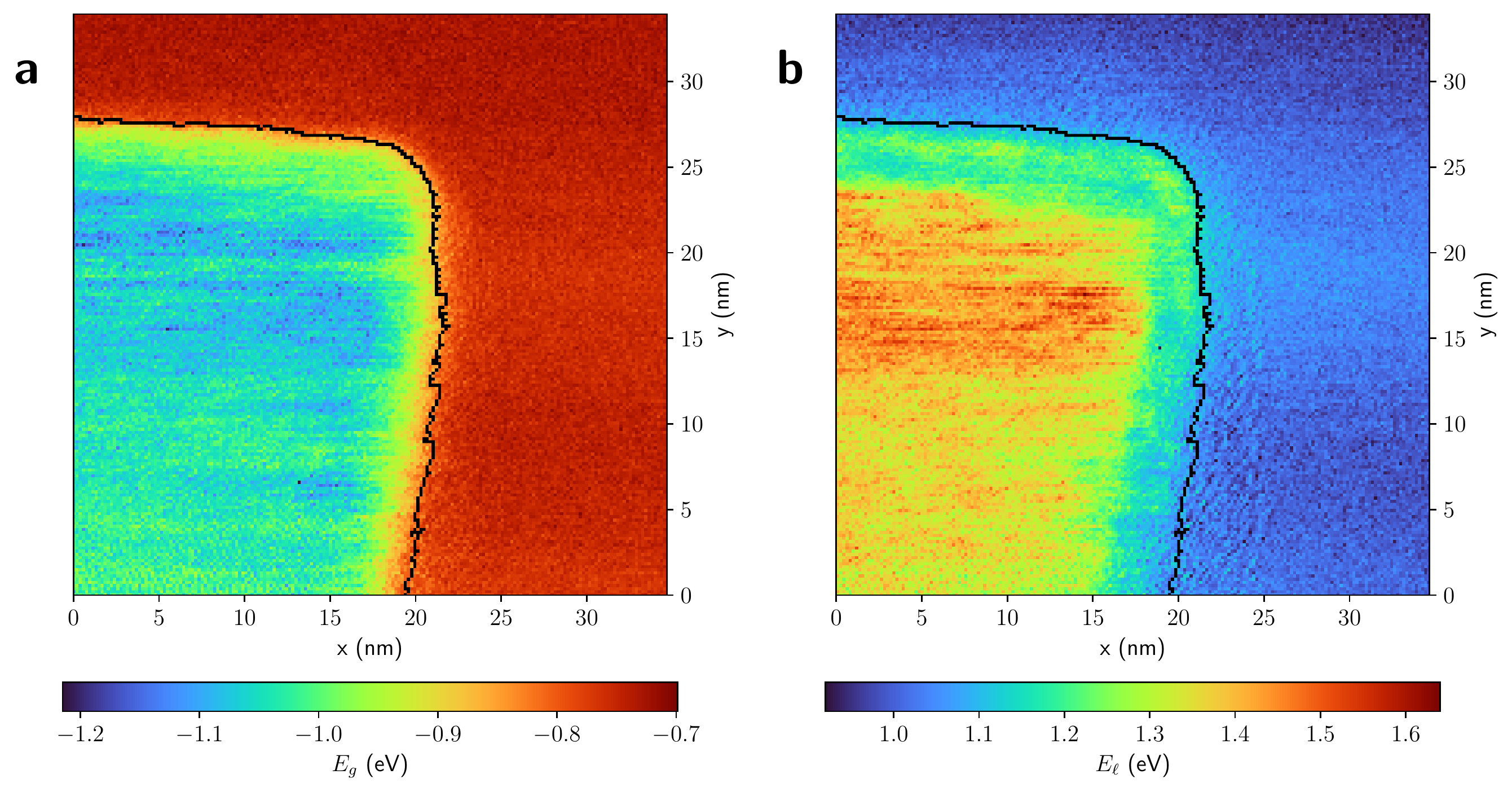}
    \caption{ \textbf{(a)} Same as Fig.~\ref{fig:subtraction_procedure}\textbf{c}. \textbf{(b)} Same as Fig.~\ref{fig:subtraction_procedure}\textbf{c} now displaying the value of $E_\ell$, the median
      of the Lorentzian peak fitted to the ZLP-subtracted energy-loss region.}
    \label{fig:peak_location}
\end{figure}

Fig.~\ref{fig:subtraction_procedure}\textbf{c} in the main manuscript
displays a spatially-resolved map with the value of $E_g$, the median
of the Lorentzian peak fitted to the ZLP-subtracted energy-gain region.
For completeness, we show in Fig.~\ref{fig:peak_location}\textbf{b} the same map
now for $E_\ell$, the median
of the Lorentzian peak fitted to the ZLP-subtracted energy-loss region.
We emphasize that our model for the energy-loss region is incomplete, given that
the data contains additional contributions beyond this dominant peak.

In the region of the specimen where the energy-gain peak exhibits the highest significance,
namely close to the edge, the value of the fitted energy-loss peak
is
$E_\ell \sim $1 eV, corresponding to approximately the energy-mirrored value
of its energy-gain counterpart.
However, the significance of this energy-loss feature is weaker,
see Fig.~\ref{fig:Bi2Te3-gain_maps_area_ratio}\textbf{d}, and only with a complete
model of the energy-loss region would it be possible to robustly disentangle
 the different energy-loss features present in the considered specimen.

\clearpage

\section{Uncertainty estimate using the Monte Carlo replica method}

\begin{figure}[htbp]
    \centering
    \includegraphics[width=.99\textwidth]{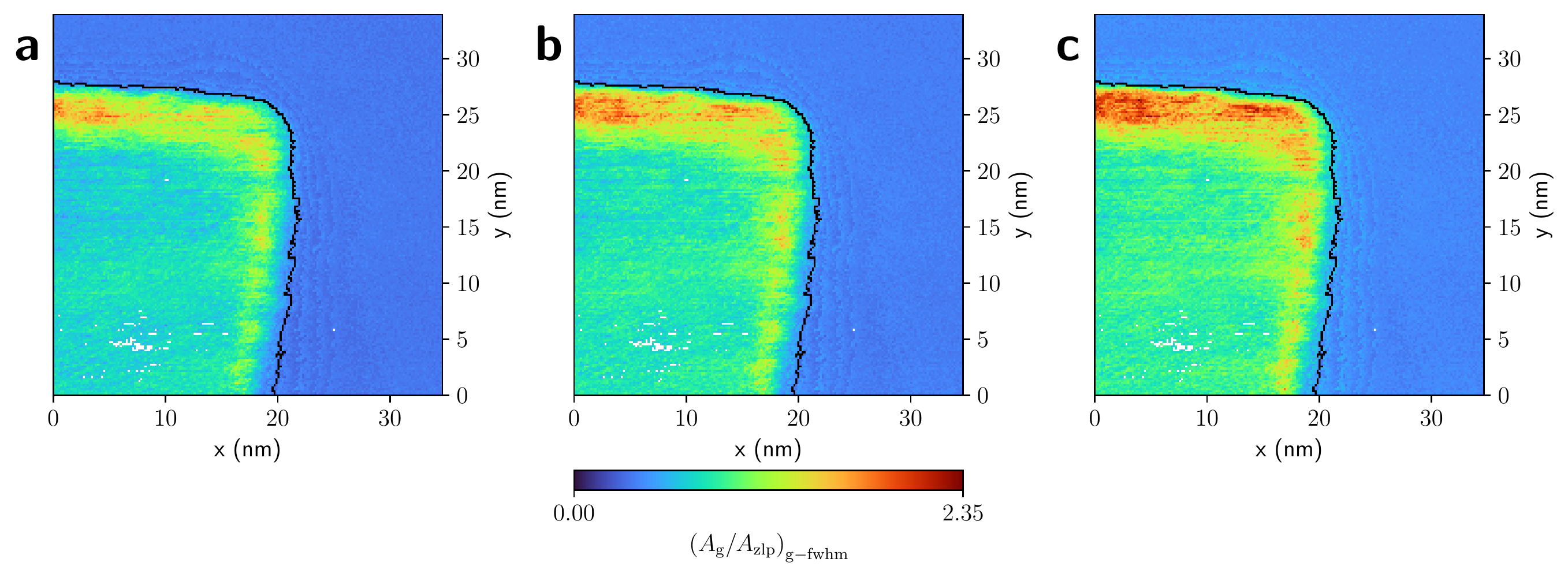}
    \caption{Spatially-resolved map of the lower 68\% CL bound (\textbf{a}), median
      (\textbf{b}), and upper 68\% CL bound (\textbf{c}) of the ratio between the area within the FWHM of the gain peak and that of the ZLP in the same $\Delta E$ region, estimated from the spread
      of the Monte Carlo replicas.
      The lower and upper bounds were already showed in  the bottom panels
      of Fig.~\ref{fig:area_ratio_uncertainty} and are repeated
      here for completeness.
    }
    \label{fig:median_and_bounds}
\end{figure}

The lower and upper 68\% confidence level intervals computed on the
ratio between the area within the FWHM of the gain peak and that of the ZLP in the same $\Delta E$ region, estimated from the spread
of the Monte Carlo replicas, were already showed in  the bottom panels
of Fig.~\ref{fig:area_ratio_uncertainty}.
In Fig.~\ref{fig:median_and_bounds}, for completeness, we also report
on the corresponding spatially-resolved map for the median of
this 68\% CL interval.
The qualitative agreement between
panels \textbf{a},\textbf{b},\textbf{c} of Fig.~\ref{fig:median_and_bounds} indicate that
uncertainties associated to the ZLP modelling and substraction procedure are moderate
and do not distort the main results obtained in this work.

\clearpage

\section{Energy-gain peaks in different $\bite{}$ specimens}
\label{sec:other_specimen}

While the results presented in the main
manuscript consider the EELS analysis of a specific
$\bite{}$ specimen, we found that
consistent results are obtained
in other comparable specimens.
In the following,
the microscope parameters that were used are the same as
the ones mentioned in the Methods section of the main manuscript.

\begin{figure}[t]
    \centering
    \includegraphics[width=\textwidth]{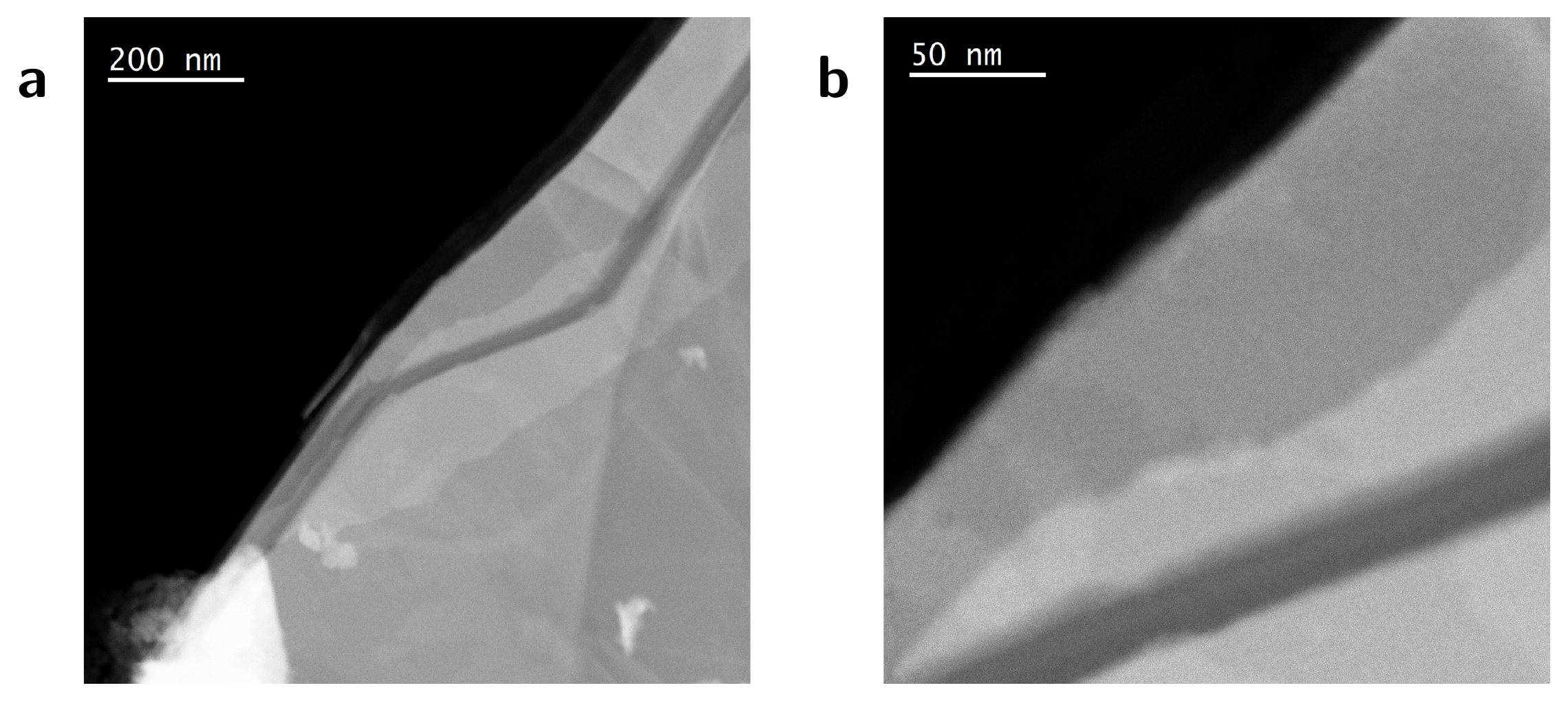}
    \caption{ \textbf{(a)} HAADF image of 
    another $\bite{}$ specimen,
    characterised by the presence
    of a channel (darker contrast).
    \textbf{(b)} A zoomed-in image of the
    same specimen, focusing on the region
    above the channel which is analysed by means of EELS. }
    \label{fig:HAADF_enhanced_region}
\end{figure}

Fig.~\ref{fig:HAADF_enhanced_region}\textbf{a} shows an HAADF image of 
    another $\bite{}$ specimen,
    different from the one in the main manuscript, and 
    characterised by the presence
    of a channel, indicated by the darker contrast.
 Fig.~\ref{fig:HAADF_enhanced_region}\textbf{b} focuses on the region
 above the channel which 
 is subsequently analysed with
     electron energy-gain spectroscopy. 
     This region is composed
     by a flat layer alongside some edges.
     The channel region
     is characterised by sharp
     variations of the specimen
     thickness, which motivate the
      inspection of the specimen
     for edge- and surface-induced
     excitations.

\begin{figure}[htbp]
    \centering
    \includegraphics[width=.99\textwidth]{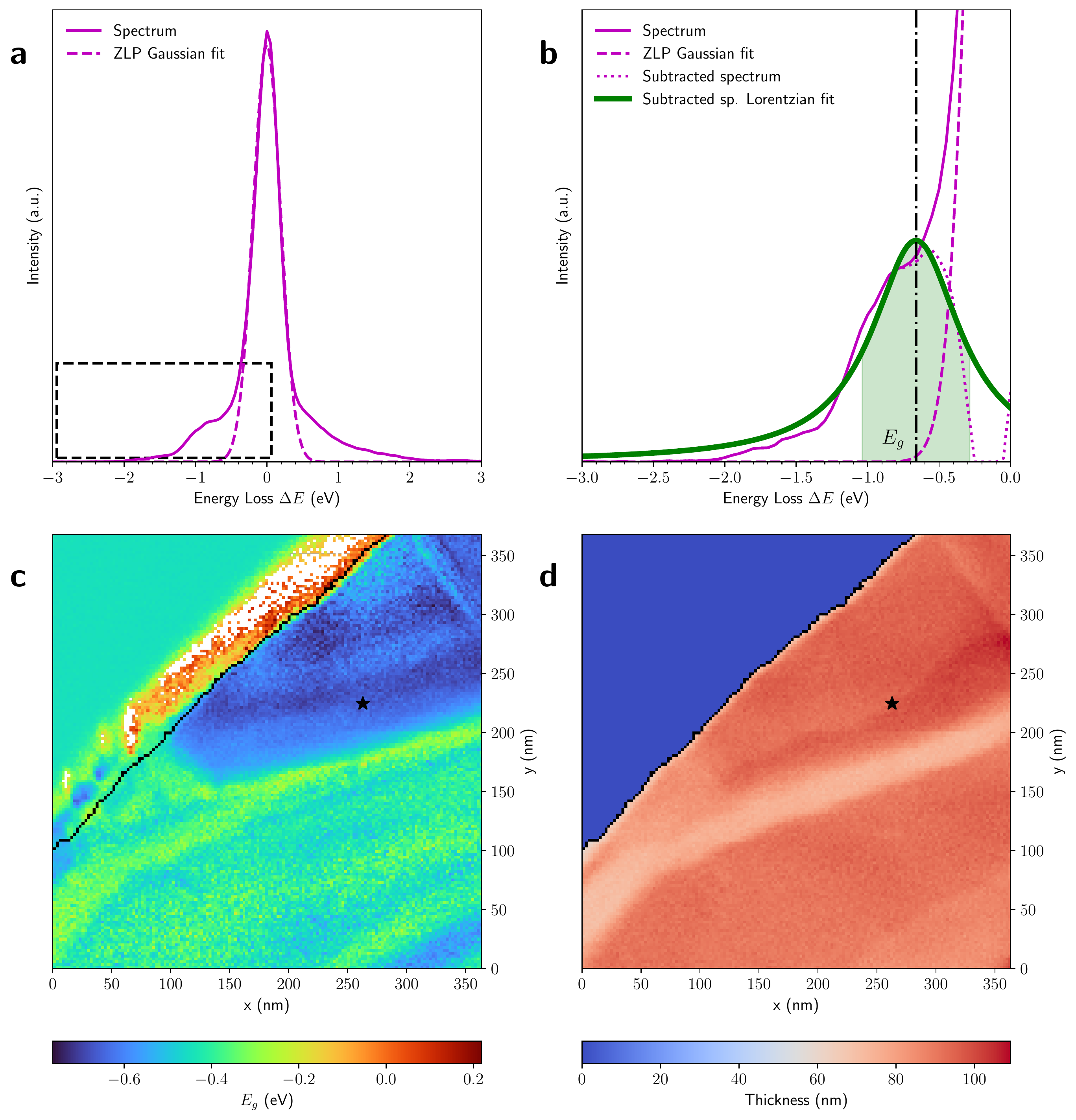}
    \caption{Same as Fig.~\ref{fig:subtraction_procedure}
    in the main manuscript for
    a different $\bite{}$ crystal.
    The star indicates the location
    in the specimen in which
    the individual EEL spectrum
    of \textbf{(a)} and \textbf{(b)}
    is extracted.
    }
\label{fig:subtraction_procedure_other_sample}
\end{figure}

\begin{figure}[htbp]
    \centering
    \includegraphics[width=.99\textwidth]{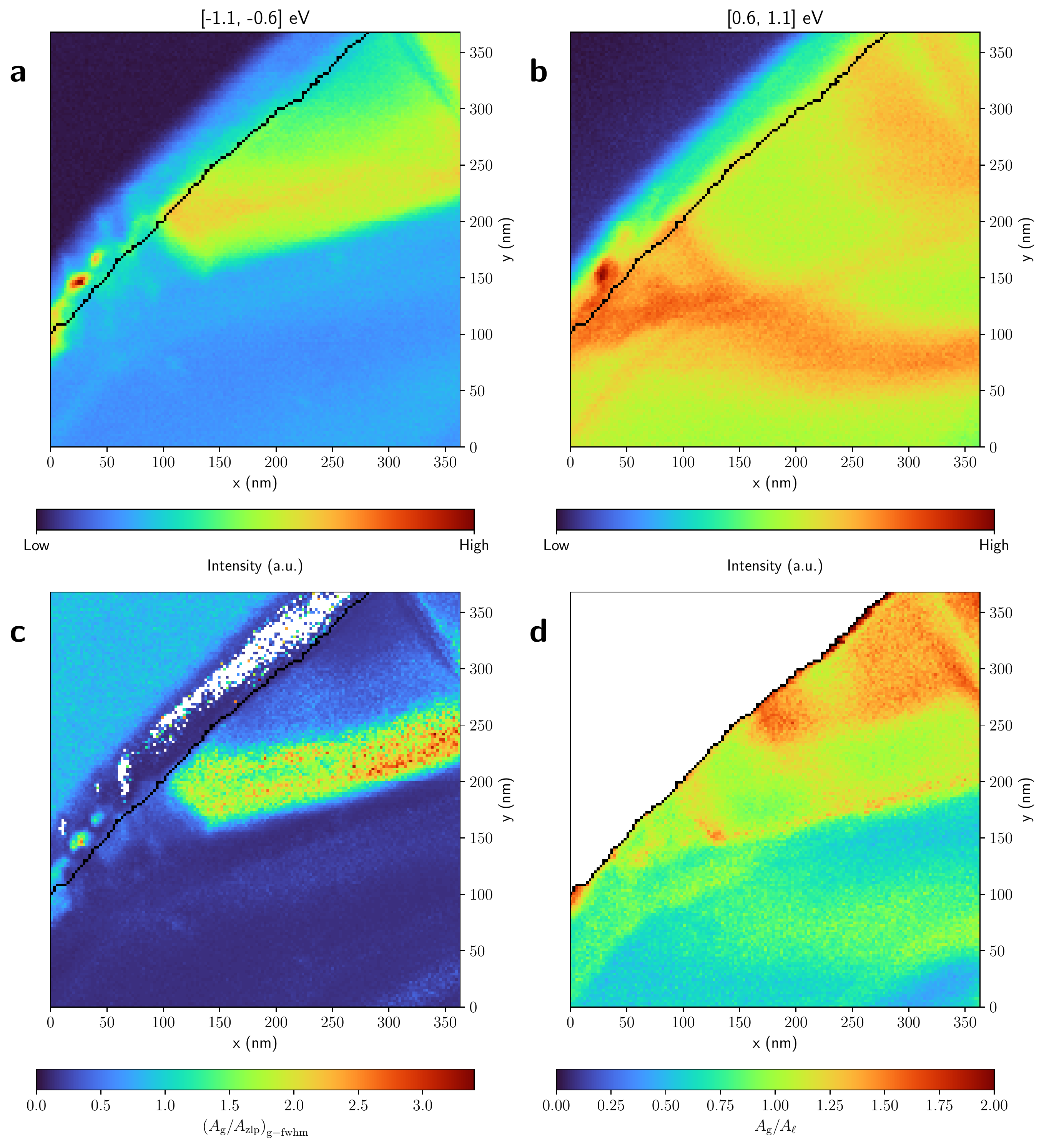}
    \caption{Same as Fig.~\ref{fig:Bi2Te3-gain_maps_area_ratio} in the
    main manuscript for
    the same $\bite{}$ crystal
    analysed in 
Fig.~\ref{fig:subtraction_procedure_other_sample}.}
    \label{fig:Bi2Te3-gain_maps_area_ratio_other_sample}
\end{figure}

Figs.~\ref{fig:subtraction_procedure_other_sample} and~\ref{fig:Bi2Te3-gain_maps_area_ratio_other_sample} 
present the same spatially-resolved
electron energy-gain analyses
of Figs.~\ref{fig:subtraction_procedure}
and~\ref{fig:Bi2Te3-gain_maps_area_ratio}
in the main manuscript respectively,
now corresponding to the $\bite{}$
specimen displayed in
Fig.~\ref{fig:HAADF_enhanced_region}.
The black line indicates the specimen boundary, and pixels in which the fitting procedure
is numerically unstable are masked out.
An enhancement of the intensity
in the $\Delta E$ gain region located
around   $-0.7$ eV
is observed in specific regions of the sample, as illustrated by the representative spectrum in Fig.~\ref{fig:subtraction_procedure_other_sample}\textbf{b} corresponding to the location
marked with a star in 
Fig.~\ref{fig:subtraction_procedure_other_sample}\textbf{a}.
This feature is most prominent
in the region above the channel
identified in Fig.~\ref{fig:HAADF_enhanced_region}
and close to the specimen surface. 
No equivalent enhancement is observed
in the loss region, where it could
be confounded by other mechanisms
contributing to energy-loss inelastic scatterings.

Applying the same peak
identification  procedure used
for the specimen in the main manuscript and further 
detailed in Sect.~\ref{sec:subtraction-procedure}, we find that in the region
where the energy gain peak
is more marked the center of the Lorentzian
is located around $\Delta E=-0.7$ eV,
see Fig.~\ref{fig:subtraction_procedure_other_sample}\textbf{d} for
the associated spatially-resolved map.
Integrating in the energy gain region defined by the window
$[-1.1,-0.6]$ eV the intensity is
highest in the region just above the channel,
Fig.~\ref{fig:Bi2Te3-gain_maps_area_ratio_other_sample}\textbf{a}.
The ratio of the area under the FWHM
of the Lorentzian peak over that
under the ZLP, Fig.~\ref{fig:Bi2Te3-gain_maps_area_ratio_other_sample}\textbf{c},
is also enhanced in the same
region, where it takes values between
$2$ and $3.5$.
In the region of the specimen with the highest signal-to-noise significance,
the position of the Lorentzian median 
is approximately constant and takes value $E_g\sim -0.7$ eV.
We verify, by means of
the same procedure adopted
for Fig.~\ref{fig:ZLP_model_comparison},
that our results are independent
of the specific choice of
functional form model for the ZLP.

The analysis presented here, performed
on a different $\bite{}$ specimen
but with the same crystal structure and
comparable features as that
of the main manuscript, confirms
the robustness of our characterisation
of the energy gain region of $\bite{}$, establishing
the presence of a distinctive gain peak
located in the region around $\Delta E =-0.7$ eV.
This feature
can be disentangled from the
dominant ZLP emission with high significance and is also found to be enhanced
in the regions of the specimen
displaying sharp thickness variations
with the associated exposed surfaces
and edges.




\end{document}